# On the structure of the Nx phase of symmetric dimers.

Anke Hoffmann[1], Alexandros G. Vanakaras[2], Alexandra Kohlmeier[3], Georg H. Mehl[3] and Demetri J. Photinos[2]

[1] Department of Inorganic and Analytical Chemistry, University of Freiburg, Albertstr. 21, 79104 Freiburg, Germany

[2] Department of Materials Science, University of Patras, Patras 26504, Greece

[3] Department of Chemistry, University of Hull, Hull, HU6 7RX, UK

## Abstract

NMR measurements on a selectively deuterated liquid crystal dimer CB-C9-CB, exhibiting two nematic phases, show that the molecules in the lower temperature nematic phase, Nx, experience a chiral environment and are ordered about a unique direction. The results are contrasted with previous interpretations of NMR measurements that suggested a twist-bend spatial variation of the director. A structural model is proposed wherein the molecules show organization into highly correlated assemblies of opposite chirality.

The classical nematic phase (N) with only orientational ordering of the molecules is theoretically and experimentally the LC phase which, due to its fundamental simplicity and technological importance, is by far the most investigated and best understood of all LC phases [1]. Thus the detection of LCs forming two nematic phases is of very high interest, as such observations pose the challenge to test and recalibrate fundamental concepts of LC phase behaviour. Direct transitions between two uniaxial thermotropic nematic phases are typically associated with the formation of more complex molecular aggregates in the low temperature nematic phase, such as column formation in the transition to the columnar nematic phase[2] of discotic or bent core materials [3]. For main chain LCPs , examples have been reported where a first order phase transition can occur between two nematic phases [4].

The recent observation of an additional, low temperature, nematic phase, termed either $N_x$, or $N_{tb}$, in very simple cyanobiphenyl based dimers with positive dielectric anisotropy, where the mesogens are separated by odd-numbered hydrocarbon spacers [5–11], and in specifically prepared difluoroterphenyl mesogens [10] with negative dielectric anisotropy, as well as in non-symmetric dimers [12], has sparked a rapidly increasing interest in this class of LCs, and particularly in the structure of this new nematic phase, $N_x$, in thin films and in the bulk. These LCs exhibit characteristic periodic stripe patterns and rope textures in thin films and an electro-optical response typically found in chiral systems, though the molecules are non-chiral[9,10,13]. XRD investigations show clearly the absence of layer reflections, confirmed by extensive miscibility calorimetric studies [5]. The presence of stripes and Bouligand arches with periodicities in the 8-10nm regime in freeze fracture TEM [11] was interpreted as formation of chiral structures on surfaces. Lastly, in a series of papers [7,8,16] focused on an extensive NMR characterization of CB-C7-CB it is argued that the interpretation of the NMR measurements is consistent with a helicoidal conical ("heliconical") nematic phase. Whilst this is in line with a spontaneous twist-bent elastic deformation predicted [14] for the nematic phases of achiral banana-shaped molecules [17], it should be stressed that the structural periodicity of 8-10 nm observed in the Nx phase of dimers [10,11] cannot be readily identified with the periodicity implied by the theoretical proposals of the "twist-bend" nematic phase [14]. Such proposals are based on the continuum theory of elasticity for the nematic phase, involving the elementary (twist, bend, splay) deformations according to the Frank-Oseen formulation of the elastic free energy. However, the validity of the fundamental assumptions of the continuum theory, concerning the slow variations of the director field [1], is problematic when the macroscopic director supposedly undergoes a reorientation of $2\theta_0$ over the distance of a half-pitch, which might be as short as one or two molecular lengths. Computer simulations, on the other hand, are not subject to such small length-scale restrictions and there it is found that bent core molecules with flexible arms [15], a model bearing structural analogies with the odd-spacer dimers, can form nematic phases consisting of chiral domains which spontaneously self-assemble into larger helical structures.

In this letter we show, based on NMR studies of a selectively deuterated material of the structure CB-C9-CB shown in Figure 1a, that the simple helicoidal conical structure is not the organisation of the $N_x$ phase in the bulk. Based on these results, we propose an organisation of the $N_x$ phase wherein the molecules form chiral domains or bundles, whose chirality is averaged out macroscopically and whose orientational order defines a unique director.

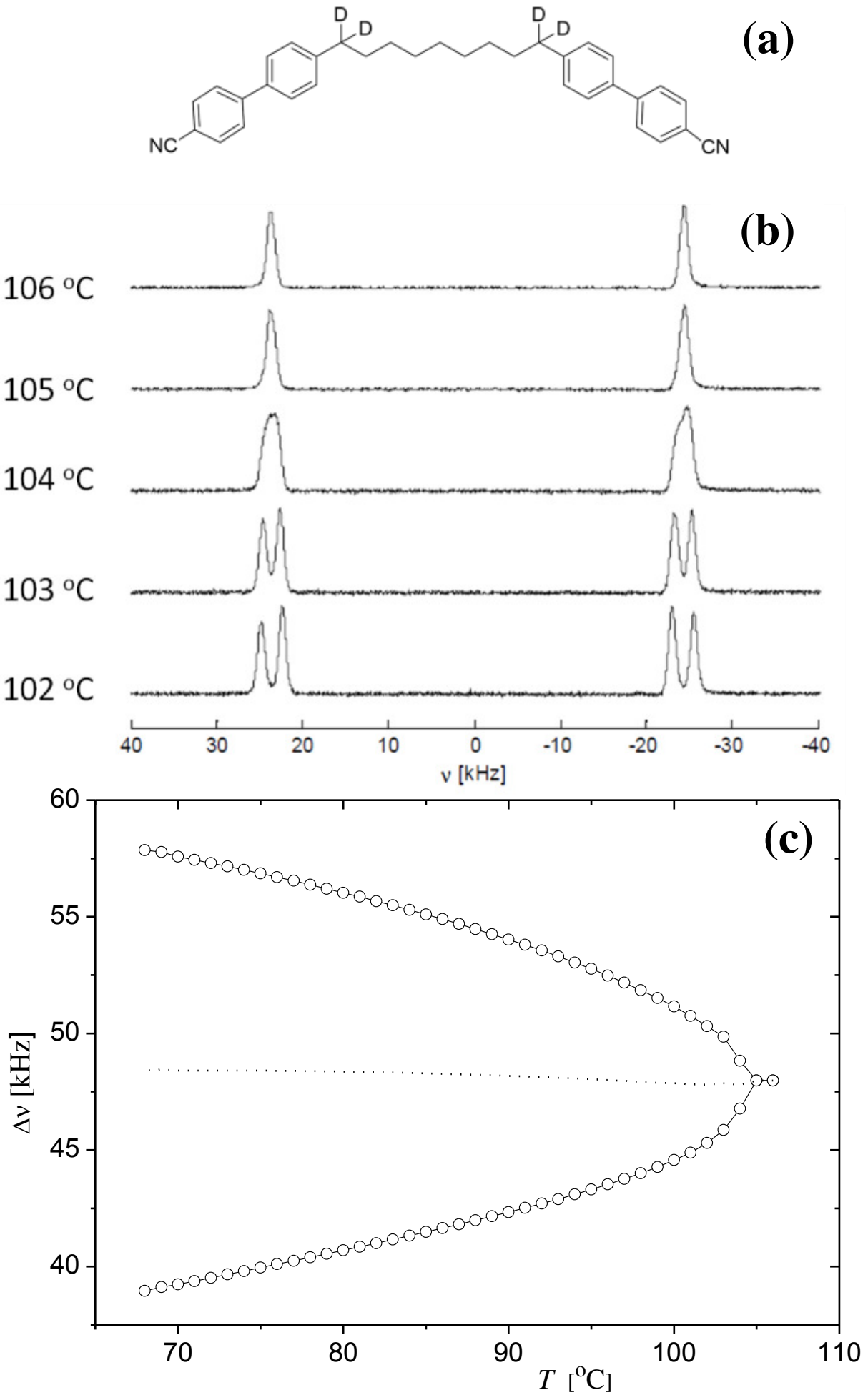

**Figure 1:** (a) Structure of the deuterated molecule CB-C9-CB, ($\alpha,\omega$-bis(4,4′-cyanobiphenyl) nonane-(1,1,9,9-$d_4$)). (b) Measured $^2$H NMR spectra recorded close to the N-Nx phase transitions (cooling run). (c) The dependence of the measured quadrupolar splittings, $\Delta\nu$, on the temperature *T* for the 1,9 deuterons. The dotted line represents the mean value of the two splitting.

The phase sequence of this statistically achiral compound is as follows. Heating: Cr 83.9 Nx 106.5 N 122.7 Iso (˚C); cooling: Iso122.1 N 106.4 Nx 48.5 Cr (˚C). The sequence is determined by DSC at 10 °C min$^{-1}$ (see ESI-4 [18]) and is close to that of non-deuterated CB-C9-CB for which the Nx phase was observed first [5] and to CB-C7-CB [7]. The materials in the N phase are typical, relatively viscous, uniaxial nematics. The onset, through a first order phase transition, of the low temperature $N_x$ phase is accompanied by a significant increase of the viscosity (see also ESI-2 [18]).

Figures 1(b, c) show the quadrupolar spectrum of aligned CB-C9-CB in the temperature range from 110-80$^o$C. The line shape and the width of the peaks suggest that the sample is uniformly oriented along the magnetic field of the spectrometer. (For experimental details see ESI-1 in [18]).

In terms of $^2$H-NMR analysis, the characteristic feature of the N-$N_x$ phase transition is the onset of the doubling of the quadrupolar peaks of the $CD_2$ sites (Figure 1(b, c)). Specifically, on entering the $N_x$ range, each of the spectral lines of the $\alpha$ -$CD_2$ groups splits into two lines with splittings $\Delta\nu_1$ and $\Delta\nu_2$, of essentially equal integrated intensity (see ESI-1.2 [18]) and with continuously increasing separation with decreasing temperature. The quantity $\overline{\Delta\nu} = (\Delta\nu_1 + \Delta\nu_2)/2$ is insensitive to temperature, with only a slight increment on cooling, see Figure 2c. While this peak-doubling alone does not allow to directly single-out a particular underlying mechanism, among several possibilities, the mechanism of loss of equivalence of the two deuterated sites [19] as a result of the onset of local chiral asymmetry in the Nx phase is strongly supported by the detailed NMR studies in [7,8] on CB-C7-CB together with other results [20] which suggest the presence of chiral asymmetry. Accordingly, our analysis and interpretation of the present data is based on adopting the hypothesis of the onset of local chiral asymmetry in the Nx phase on a time scale larger than the time-averaging involved in the NMR measurement. Within this interpretation, the intensities of the two sub-peaks should be strictly equal as they correspond to equal numbers of (orientationally inequivalent) deuterated sites.

The measured splitting in the quadrupolar spectrum of the α-$CD_2$ group in the nematic phase is related to the respective orientational order parameter $S_{\alpha-CD} = \left\langle P_2(\hat{\mathbf{e}}_{\alpha-CD} \cdot \hat{\mathbf{H}}) \right\rangle$ according to $\Delta\nu = -\frac{3}{2} q_{CD} S_{\alpha-CD}$ , with $\hat{\mathbf{e}}_{\alpha-CD}$ denoting the orientation of the deuterated CD bond, $\hat{\mathbf{H}}$ the direction of the spectrometer magnetic field, $q_{CD}$ the quadrupolar coupling constant (here set at the value of $168\,kHz$), $P_2$ the second Legendre polynomial and the angular brackets indicate ensemble time-averaging [21]. From the quadrupolar spectrum of the α-$CD_2$ group in the high temperature *N* phase (Figure 3a) at 106°C, with a splitting of $\Delta\nu \simeq 48\,kHz$, we obtain the value $-0.19$. for $S_{\alpha-CD}$. Assuming a fixed tetrahedral angle of the a-CD bonds relative to the para-axis of the cyanobiphenyl mesogenic core, we obtain for the order parameter $S_{core}$ of the mesogenic unit the value 0.57; this falls within the typical range for nematics.

The crucial NMR experiment for testing the formation of the helicoidal conical structure in the bulk makes use of the high viscosity of the Nx phase. This allows collection of NMR spectra with the sample oriented parallel as well as perpendicular to the magnetic field. Specifically, when rotating the initially aligned sample to 90° with respect to the magnetic field, there is sufficient time to measure the NMR spectrum in this configuration before the magnetic field reorients the sample. Figure 2a shows the quadrupolar spectrum at 90 °C before (red) and after (blue) a rotation of the sample by 90° with respect to the magnetic field. Notably, the spectrum of the flipped sample, when measured 10s after the flip (blue in Figure 2b), is essentially identical to the initial spectrum (red), indicating that in this time interval the sample relaxes back along the direction of **H**. From the spectra in Figure 2a we observe that the 90° flipped sample (i) does not exhibit any line broadening, the line shape being the same as the initially aligned sample, in fact with some slight narrowing and (ii) the quadrupolar peaks are at half the initial splitting. Moreover, the spinning sample patterns of Figure 2c are typical of a cylindrical distribution of the director about the axis of spinning, indicating that the sample is uniaxial.

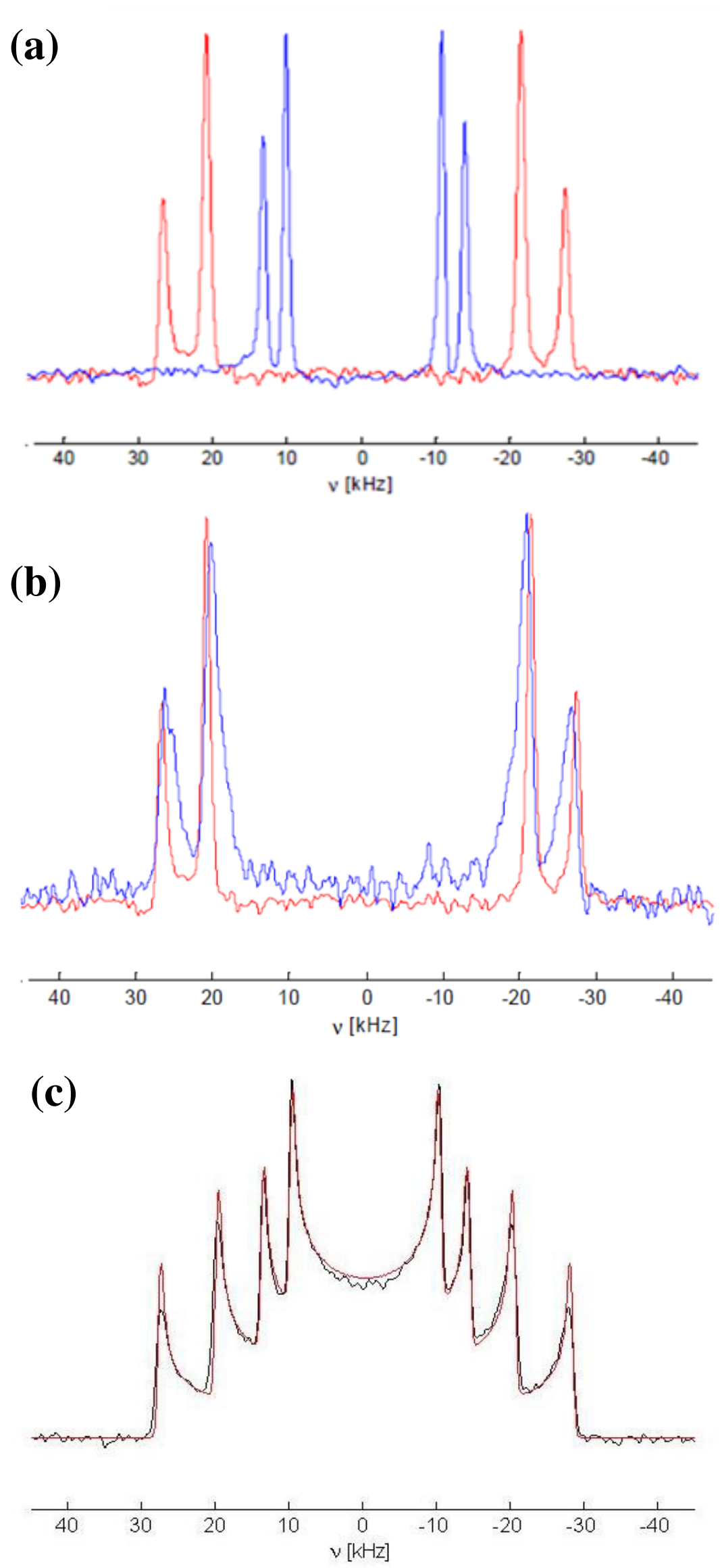

**Figure 2. (a) $^2$H** NMR spectra of the aligned sample (red) and after a flip by 90° with respect to the magnetic field (Blue) in the Nx phase at 90$^o$C. (b) $^2$H NMR spectrum 10s after the 90° flip (blue); the red line corresponds to the initial spectrum (in the Nx phase at 90$^o$C).(c) Spinning sample (100rpm) $^2$H NMR spectrum in the Nx phase at 80$^o$C.

These results have a direct impact on the structures possible for the $N_x$ phase. An immediate implication is the dismissal of the simple heliconical structure: The configuration of the nematic director for a heliconical structure is given in Figure 3a. In this model the director follows a configuration described by $\mathbf{n} = (\sin\theta_0 \cos kZ, \sin\theta_0 \sin kZ, \cos\theta_0)$, about an axis identified with the Z axis of the phase, which, in turn, aligns parallel to the external magnetic field and therefore the director makes a fixed angle $\theta_0$ with the field; here $k = 2\pi / q$ and $q$ is the pitch of the helix. In such a configuration of the director, rotating the sample by 90° about an axis perpendicular to the magnetic field generates a distribution of angles, $\theta_{n,H}$, in the range $\pi/2-\theta_0 < \theta_{n,H} < \pi/2+\theta_0$, whose breadth increases with increasing $\theta_0$. However, the measured spectra of Figure 2a show no such distribution. Instead, a spectrum typical of the rotation of a single director, from parallel to perpendicular orientation relative to the magnetic field, is seen. No such spectra should be obtained if the director had the twist-bend helical configuration with any appreciable value (5° or larger, see ESI-3.1 [18]) of the "tilt" angle $\theta_0$. Thus, the 90°-rotated and the spinning sample spectra indicate unambiguously that the twist-bend helical configuration of the director, either throughout the sample or in domains of opposite handedness, is not obtained in the $N_x$ phase under the action of the spectrometer magnetic field. In contrast, the data suggest a picture wherein the macroscopic sample consists of microscopic domains which are chiral. Each such domain exhibits molecular orientational ordering along a unique direction (the local director of the domain) which aligns parallel to the magnetic field, as shown schematically in Figure 3c.

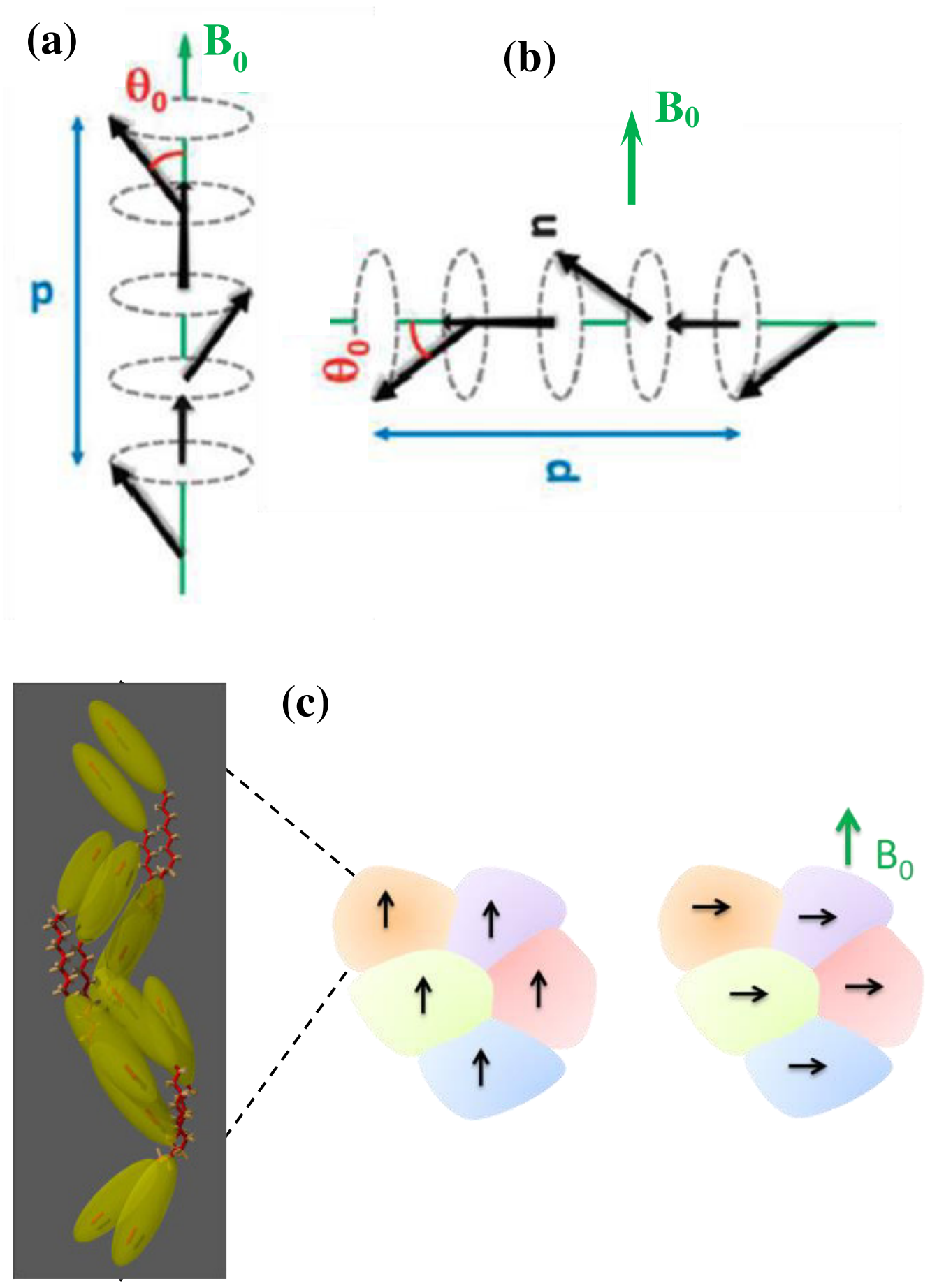

**Figure 3.** Top: Schematic representation of the hypothetical (not supported by the present findings) heliconical configuration of the nematic director for (a) the magnetically aligned sample [16] and (b) the sample rotated by 90° relative to the magnetic field. The latter configuration is directly disproved by the NMR measurement while there is no direct indication for the former from the aligned sample measurement. Bottom: Proposed chiral domain structure of the Nx phase. The inset (left) shows a possible disposition of the mesogenic units within each domain, stressing the lack of full alignment of all the units parallel to a unique direction. The internal orientational order of each domain is represented by a local director. The magnetic field aligns the local directors of the domains (middle). On rotating the sample (right), the mutual alignment of the domains persists sufficiently long on the NMR measurement time-scale.

Clearly, there is no helical, nor heliconical, molecular ordering within the chiral domains. Instead, the dimer molecules within these domains assume predominantly twisted, chiral, conformations [21] and the enantio-selective local molecular packing confers to the domain its chirality. As the molecules are statistically achiral, each chiral conformation is macroscopically equiprobable with its mirror image. Thus there are equal amounts of right and left handed chiral domains. Furthermore, the complete separation of each peak into two sub-peaks (there is marginal intensity between the two sub-peaks), exhibited by the measured spectra toward the low temperature end of the $N_X$ phase (see figure 1(b)), suggests that the sample consists entirely of chiral domains, with practically no achiral regions which could be detected within the experimental resolution of the measurements. At higher temperatures in the $N_X$ phase, the sub-peaks are overlapping and there it cannot be excluded that part of the intensity could correspond to a substantial achiral contribution whose extent, however, cannot be directly estimated from the spectra. The chiral domain picture is in accord with a periodicity of 9nm of the twist found for crystalline samples [22] and in very similar systems, where TEM data indicates periodicities of 8 nm [11]. It is also consistent with the dramatic increase of the viscosity on going from the N to the Nx phase, as the reorientation processes in the latter would involve the rotation of entire domains of strongly correlated molecules.

The NMR measurements show that, within the entire range of the Nx phase, the ordering of the mesogenic units along the magnetic field remains relatively low (0.5 to 0.6); this is in line with the chiral domain picture and no heliconical configuration needs to be invoked (see also ESI-3.2 [18]). Specifically, the mesogenic units of the dimers, due to the statistical dominance of nonlinear conformations, cannot attain a simultaneous alignment along any given direction (see Figure 3c, left). Accordingly, though a direction of maximal alignment (i.e. the local director) can be uniquely defined, within each domain and for the entire aligned sample, the respective value of the order parameter is well below the value of 1. This is also the case for the crystal phase of the CB-C9-CB dimers where [22] the molecules are packed in twisted conformations and the order parameter of the mesogenic units is significantly lower than 1. Hence, the structure of the $N_x$ phase for the investigated system is best described as a chiral domain ($N_{CD}$) structure which can be uniformly aligned by the magnetic field to produce a unique nematic director, common for all the domains. It is stressed, however, that the results presented here do not preclude the possibility that, in the absence of the aligning magnetic field, the chiral domains could produce, via soft self-assembly, helical structures on a larger length scale under particular surface anchoring.

**References.**

[1] P. G. de Gennes and J. Prost, *The Physics of Liquid Crystals* (Oxford University Press, 1995).

[2] P. H. J. Kouwer, W. F. Jager, W. J. Mijs, and S. J. Picken, Macromolecules **33**, 4336 (2000).

[3] C. Keith, A. Lehmann, U. Baumeister, M. Prehm, and C. Tschierske, Soft Matter **6**, 1704 (2010).

[4] G. Ungar, V. Percec, and M. Zuber, Polym. Bull. **32**, 325 (1994).

[5] C. S. P. Tripathi, P. Losada-Pérez, C. Glorieux, A. Kohlmeier, M.-G. Tamba, G. H. Mehl, and J. Leys, Phys. Rev. E **84**, 041707 (2011).

[6] P. A. Henderson and C. T. Imrie, Liq. Cryst. **38**, 1407 (2011).

[7] M. Cestari, S. Diez-Berart, D. A. Dunmur, A. Ferrarini, M. R. de la Fuente, D. J. B. Jackson, D. O. Lopez, G. R. Luckhurst, M. A. Perez-Jubindo, R. M. Richardson, J. Salud, B. A. Timimi, and H. Zimmermann, Phys. Rev. E **84**, 031704 (2011).

[8] L. Beguin, J. W. Emsley, M. Lelli, A. Lesage, G. R. Luckhurst, B. A. Timimi, and H. Zimmermann, J. Phys. Chem. B **116**, 7940 (2012).

[9] V. P. Panov, R. Balachandran, J. K. Vij, M. G. Tamba, A. Kohlmeier, and G. H. Mehl, Appl. Phys. Lett. **101**, 234106 (2012).

[10] V. Borshch, Y.-K. Kim, J. Xiang, M. Gao, A. Jákli, V. P. Panov, J. K. Vij, C. T. Imrie, M. G. Tamba, G. H. Mehl, and O. D. Lavrentovich, Nat. Commun. **4**, 2635 (2013).

[11] D. Chen, J. H. Porada, J. B. Hooper, A. Klittnick, Y. Shen, M. R. Tuchband, E. Korblova, D. Bedrov, D. M. Walba, M. A. Glaser, J. E. Maclennan, and N. A. Clark, Proc. Natl. Acad. Sci. **110**, 15931 (2013).

[12] M. G. Tamba, A. Kohlmeier, and G. H. Mehl, in *12th ECLC Book Abstr.* (Rhodes, Greece, 2013), p. O12.

[13] V. P. Panov, M. Nagaraj, J. K. Vij, Y. P. Panarin, A. Kohlmeier, M. G. Tamba, R. A. Lewis, and G. H. Mehl, Phys. Rev. Lett. **105**, 167801 (2010).

[14] I. Dozov, EPL Europhys. Lett. **56**, 247 (2001).
[15] S. D. Peroukidis, A. G. Vanakaras, and D. J. Photinos, Phys. Rev. E **84**, 010702(R) (2011).
[16] C. Greco, G. R. Luckhurst, and A. Ferrarini, Phys. Chem. Chem. Phys. **15**, 14961 (2013).
[17] D. Chen, M. Nakata, R. Shao, M. R. Tuchband, M. Shuai, U. Baumeister, W. Weissflog, D. M. Walba, M. A. Glaser, J. E. Maclennan, and N. A. Clark, ArXiv13083526 Cond-Mat (2013).
[18] Supporting Information
[19] K. Czarniecka and E. T. Samulski, Mol. Cryst. Liq. Cryst. **63**, 205 (1981).
[20] D. O. López, N. Sebastian, M. R. de la Fuente, J. C. Martínez-García, J. Salud, M. A. Pérez-Jubindo, S. Diez-Berart, D. A. Dunmur, and G. R. Luckhurst, J. Chem. Phys. **137**, 034502 (2012).
[21] D. J. Photinos, E. T. Samulski, and H. Toriumi, J. Chem. Soc. Faraday Trans. **88**, 1875 (1992).
[22] K. Hori, M. Iimuro, A. Nakao, and H. Toriumi, J. Mol. Struct. **699**, 23 (2004).

# Supporting Information

## On the structure of the Nx phase of symmetric dimers.

Anke Hoffmann[1], Alexandros G. Vanakaras[2], Alexandra Kohlmeier[3],

Georg H. Mehl[3], Demetri J. Photinos[2]

[1] Department of Inorganic and Analytical Chemistry, University of Freiburg, 79104 Freiburg, Germany

[2] Department of Materials Science, University of Patras, Patras 26504, Greece

[3] Department of Chemistry, University of Hull, Hull, HU6 7RX, UK

*This supporting information file contains details on the NMR measurement and analysis of the measured spectra, a presentation of the influence of rotational viscosity on the NMR spectra and a discussion of certain additional difficulties with interpretations based on the twist-bend model and the associated proposal of a heliconical configuration of the director. The DSC of the studied compound is also included.*

## 1. Details of NMR measurements

**1.1 Experiments.** All spectra were measured on a Bruker Avance 500 NMR spectrometer ($B_0$=11.7 *T*, $\nu_{2H}$ = 76.8 *Hz*). Solid echo experiments were performed with 90° pulse length of 4 *μs* and a pulse spacing of 43 *μs*. For the rotation of the sample a homebuilt probehead with a servomotor was used, triggered from the NMR console (Figure S1a). The sample can be rotated around a fixed axis perpendicular to the magnetic field. Samples were prepared by filling the substance into a 2 cm long glass tube with 4 mm diameter and sealed with a teflon/silicon plug (punched out from a septum). This tube was then fixed in the middle of a 5mm glass tube making use of teflon band (Figure S1b). The flipping experiments were performed by flipping the sample by 90° around an axis perpendicular to the static magnetic field, data acquisition, and then flipping the sample back to the initial position for the recycle delay (1s). The acquisition starts (earliest) 200 ms after triggering the sample flip, which is the time at which the sample has definitely reached the new orientation. This has been tested extensively with standard samples to get the proper timing and to verify the precision of the angular setting and test the fixation of the sample. Also, this probe has been in use for several years and has been proven to give reliable results for several low molecular weight as well as polymeric liquid crystals.

The spinning experiment spectra shown in Fig 2c. were recorded with 100 *rpm* spinning rate about an axis perpendicular to the static magnetic field. Spectra recorded with 6, 60 and 200 *rpm* showed no significant differences at the given temperature.

Cooling and heating sequences were run with temperature steps of 1K. After each step the sample was equilibrated for 10 minutes before the start of the acquisition. For the experiments under $90^o$ flipping and under continuous spinning the sample was heated to the isotropic phase and then slowly cooled to the desired temperature. The splittings obtained from different cooling runs agree within an error of less than 1%.

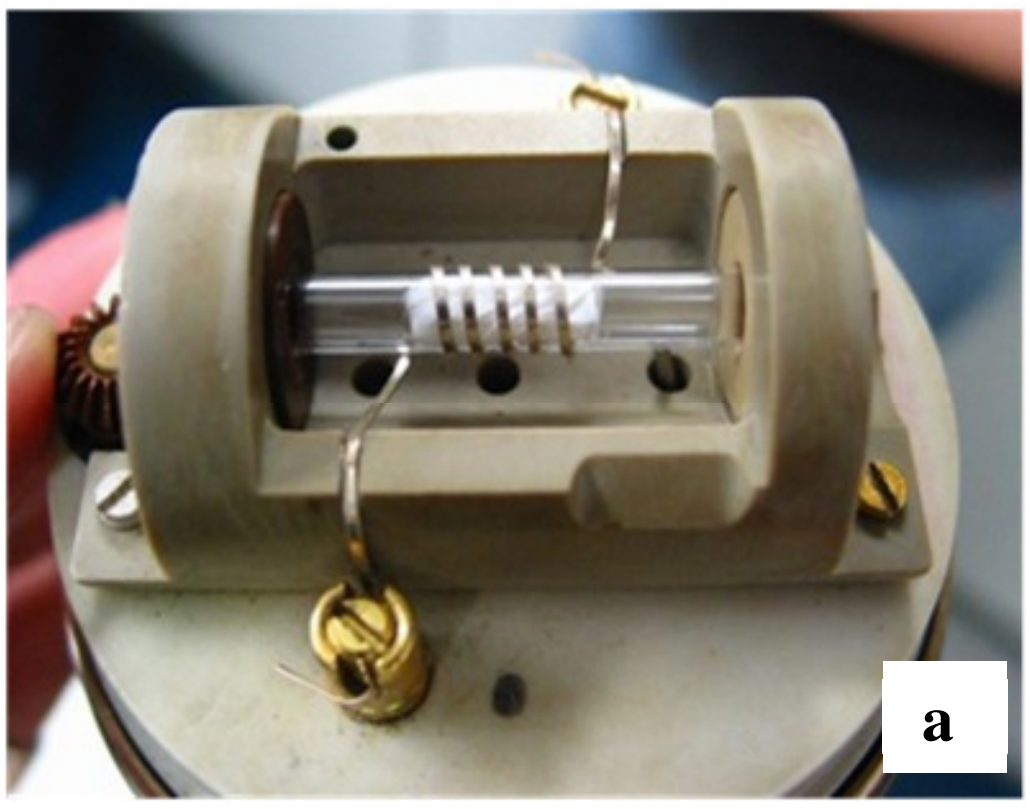

**Figure S1** a.) Probe with a mounted sample tube (not CB-9-CB). Cog-wheel for rotation can be seen on the left.
b.) Sample tube containing CB-9-CB.

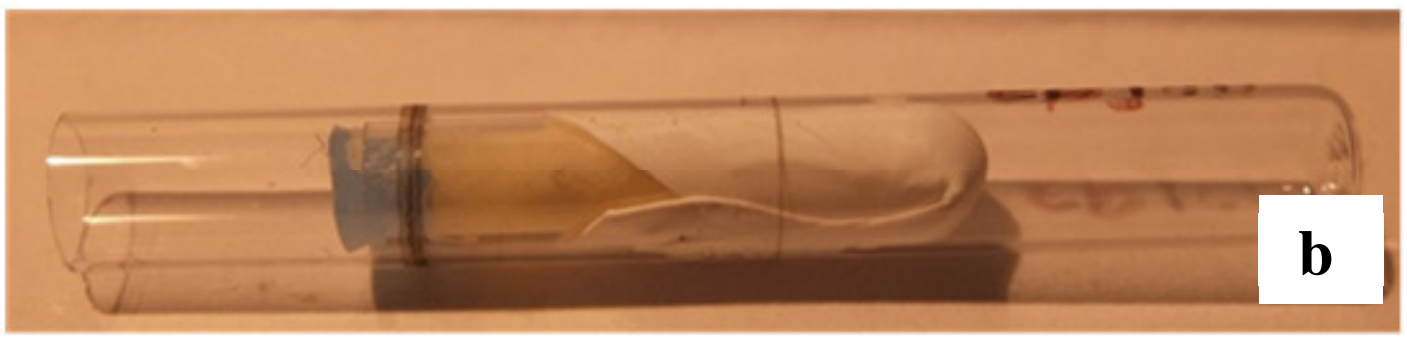

## 1.2 Analysis of measured spectra.

***(a) Aligned sample spectra.***

The doubling of each of the quadrupolar peaks of the α-$CD_2$ sites in the $N_x$ phase produces two sub-peaks of different heights and widths. Specifically, the height of the outer sub-peak is appreciably lower but its width is larger than that of the inner sub-peak. The integrated intensity, as it comes out from the direct manual integration in the NMR software is about 10% smaller for the outer peaks. However, this neglects the fact that a part of the outer peaks (the asymmetric 'foot' to the middle) lies underneath the inner peaks. When this is taken into account the relative intensities are readily shifted to equality, within the experimental resolution. Furthermore, the simulation of the spectra recorded under continuous rotation, discussed in the next section, is based on summing the two sub-peaks with equal integrated intensities, and this yields the essentially exact agreement with the measured spectral distribution of intensity.

***(b) Spinning sample spectra.***

The experimental spectra obtained under continuous spinning of the sample about an axis perpendicular to the magnetic field were simulated by a powder sum for cylindrical distribution of the nematic directors (i.e. without a scaling factor sin(β) as for spherical distribution), namely

$$I(t)=\sum_{i=1}^{N}\cos(2\pi\delta_Q(3\cos^2(\beta_i)-1)t)$$

Where $\delta_Q$ is the anisotropy of the (time averaged) quadrupolar coupling tensor; the asymmetry of the tensor is neglected. A set of 200 polar angles $\beta$, equally distributed in the range of 0 to $\pi$, was used (i.e. $N$=200; a simulation with $N$=400 showed no significant differences). The spectrum shown in figure 2(c) was simulated as the sum of two cylindrical powder patterns with anisotropies $\delta_Q$ of 28 *kHz* and 20.2 *kHz* – the values of the quadrupolar splitting obtained from the temperature run (Fig 1c). The patterns were added without additional weighting factors (i.e. the two patterns have the same integrated intensity) and scaled to the height of the inner peaks of the experimental spectrum.

***(c) Simulation of the aligned and 90$^o$ –flipped spectra in the N$_X$ phase.***

In figure S2 we present the experimental NMR spectra of the aligned (red) and of the flipped by 90$^o$ with respect to magnetic field samples, (see also fig Fig 2(a) of the paper). Best fitting curves are also shown for both sets of measured spectra. The line shapes, $L_{0(90)}(\nu)$ , of these spectra are fitted assuming a superposition of two Lorentzians $\Lambda(\nu;\delta\nu,w)=w/\left(\pi(w^2+(\nu-\delta\nu)^2\right)$ of the *same integrated intensity*:

$$L_{\theta_{H,n}}(\nu)=\sum_{r=\pm 1}\Lambda\left(\nu;r\frac{\delta\nu_Q(\theta_{H,n})}{2},w_0(\theta_{H,n})\right)+\Lambda\left(\nu;r\frac{\delta\nu_Q'(\theta_{H,n})}{2},w_0'(\theta_{H,n})\right) \quad \text{(S.1)}$$

Here, $\theta_{H,n}=0$ or 90, denotes the angle between the magnetic field and the director for the aligned and the 90$^o$-flipped samples, respectively.

| **Table I.** Best-fit parameters for the simulated spectra with eq **Error! Reference source not found.**. | | | | |
|---|---|---|---|---|
| $\theta_{H,n}$ | $\delta\nu_Q(\theta_{H,n})/2$ | $w_0(\theta_{H,n})$ | $\delta\nu'_Q(\theta_{H,n})/2$ | $w'_0(\theta_{H,n})$ |
| $0^0$ | 26.958 | 0.734 | 21.134 | 0.380 |
| $90^0$ | 13.545 | 0.374 | 10.486 | 0.260 |

We note here that for both peaks we have, $\delta\nu(90) = \delta\nu(0)/2$ and $\delta\nu'(90) = \delta\nu'(0)/2$, as expected for a uniaxial nematic.

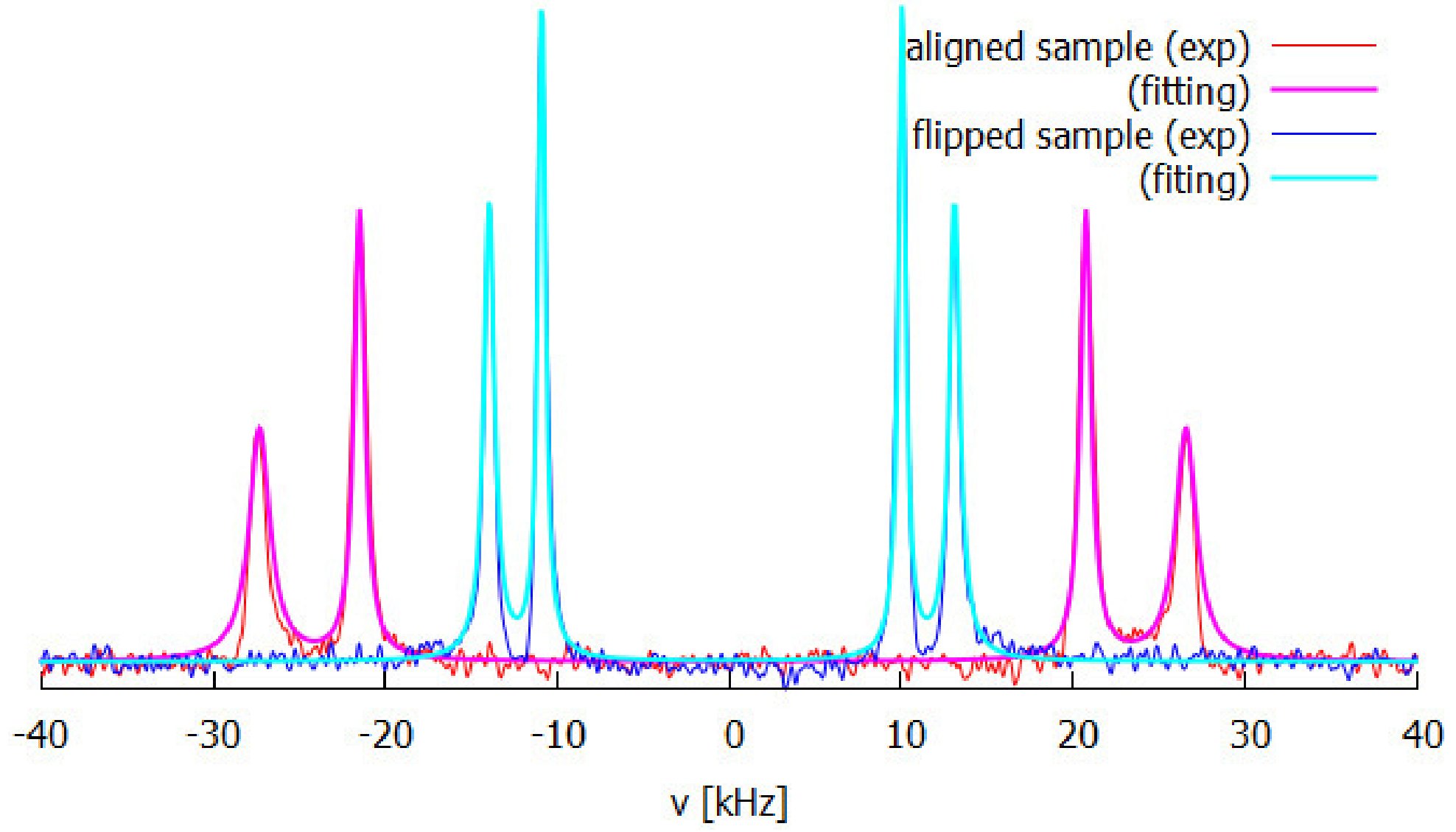

**Figure S2:** $^2$H NMR spectra of the aligned sample (red) and after a flip by 90° with respect to the magnetic field (blue) in the Nx phase at 90°C. The thick lines (magenta for the aligned sample and cyan for the 90°-flipped) are simulated spectra assuming Lorentzian line-shapes according to eq **Error! Reference source not found.**, with the best-fit parameter values listed in Table I.

## 2. The effects of viscosity on the 90°-rotated and spinning sample spectra.

At 80°C, spectra recorded under continuous spinning (Figure 2(c)) showed no significant difference when recorded with spinning rates of 6 *rpm*, 60*rpm*, 100 *rpm* and 200 *rpm*. This independence on the spinning rate indicates that a uniform cylindrical distribution of the nematic director is generated by the spinning of the sample and that such distribution is not influenced by hydrodynamic effects, as these would show a dependence on the spinning-rate. Furthermore, the spectra had not relaxed back to equilibrium two hours after the rotation was stopped. Such slow relaxation rates, together with the absence of significant hydrodynamic effects on spinning, are direct implications of the high viscosity of the $N_X$ phase.

The orientational relaxation times show a relatively rapid variation with temperature within the $N_X$ phase and a dramatic drop on entering the N phase. Thus, the spectra shown in Figure 2(b), recorded at the temperature of 90°C, are far from being fully relaxed 10s after the 90° flip. A respective spectrum recorded at 100°C is only partially relaxed after 4s. At 104 °C, the spectra recorded under continuous spinning look different for 1*rpm*, 6*rpm*, 60*rpm*, indicating the onset of appreciable hydrodynamic effects on the director distribution. Lastly, at 106°C, i.e. near the low temperature end of the N phase range, the spectra before and after the 90° flip are identical, i.e. the orientation fully relaxes within less than 200 *ms*. Spectra recorded under continuous spinning still show a single splitting for rotation up to 2000 *rpm*; only for rotations of 4000 *rpm* and 5000 *rpm* the spectra get broad. Accordingly, the orientational relaxation time (~$10^{-3}$sec) at this temperature in the N phase is four orders of magnitude faster than the respective time (~10 sec) at the temperature of 90°C in the $N_X$ phase.

## 3. Further difficulties with the heliconical interpretation.

As detailed in the main text the primary inconsistencies regarding the interpretation of the experimental data with the heliconical model of the director configuration are (i) the measured 90°-rotated and spinning sample NMR spectra indicate rather directly that the entire sample exhibits a single, uniform, director, and (ii) the structural periodicity of 10nm sets a rather short length scale on which not only the description of a continuous twist and/or bend spatial variation of the director would be invalid, but even the validity of the definition of a nematic director would be questionable. Notably, chiral domains in the nematic phase of achiral compounds exhibiting dimerisation through hydrogen bonding were reported more that 15 years ago[1]; however, such domains were found to exhibit macroscopic twist deformation of the director in the optical regime.

In addition to the above deficiencies, the heliconcal interpretation is in difficulty with the consistent reproduction of the orientational ordering implied by the NMR measurements and also with the slow orientational relaxation observed directly in the $N_X$ phase, as described in section 2. These additional difficulties are discussed separately below.

[1] S.I. Torgova, L. Komitov and A. Strigazzi, Liquid Crystals, **24**(1), 131-141 (1998).

### 3.1 Spectral line-shape for a hypothetical heliconical configuration with the magnetic field perpendicular to the helix axis

Here we present the calculation of the NMR spectrum of a sample assuming that the $N_X$ phase is a twist-bent nematic presenting a heliconical distribution of the director field. The calculated spectra are then contrasted with the measured spectra.

As indicated in the main text of the paper, the splitting associated with an α-C-D bond, is given by the ensemble average:

$$\delta\nu_Q = \nu_Q \left\langle \frac{3}{2}(\hat{H}\cdot\hat{e}_{\alpha-CD})^2 - \frac{1}{2} \right\rangle \tag{S.2}$$

Where $\hat{H}$ denotes the direction of the magnetic field and $\hat{e}_{\alpha-CD}$ the direction of the C-D bond. In a uniaxial nematic phase, with the director denoted by $\hat{n}$, the splitting can be expressed in terms of the "bond order parameter"

$$S_{C-D} \equiv \left\langle \frac{3}{2}(\hat{n}\cdot\hat{e}_{\alpha-CD})^2 - \frac{1}{2} \right\rangle \tag{S.3}$$

and the orientation of the director relative to the magnetic field as follows:

$$\delta\nu_Q = \nu_Q S_{C-D}\left( \frac{3}{2}(\hat{H}\cdot\hat{n})^2 - \frac{1}{2} \right) \tag{S.4}$$

With the help of equations (S.1)-(S.4) we can calculate the spectral line-shape for a hypothetical heliconical configuration with the magnetic field perpendicular to the helix axis, obtained by 90-flip of the aligned sample about an axis perpendicular to the magnetic field: In a macroscopic frame where the Z axis is identified with the helix axis and the direction of the magnetic field (perpendicular to the helix axis) is identified with the Y axis, the assumed heliconical distribution of the director, $\hat{n} = (\sin\theta_0 \cos kZ, \sin\theta_0 \sin kZ, \cos\theta_0)$, would imply the following distribution of the directional term in the rhs of eq (S.4) over the angle $\varphi = kZ$,

$$\frac{3}{2}(\hat{H}\cdot\hat{n})^2 - \frac{1}{2} = \frac{3}{2}(\sin\theta_0 \sin\varphi)^2 - \frac{1}{2} \tag{S.5}$$

In this case we have, according to eq(S.4) a $\varphi$-distribution of splittings,

$$\delta\nu_Q(\theta_0,\varphi) = \frac{\nu_Q S_{C-D}}{2}\left(3(\sin\theta_0 \sin\varphi)^2 - 1\right) = \frac{\delta\nu_Q(0)}{2}\left(3(\sin\theta_0 \sin\varphi)^2 - 1\right) \quad \text{(S.6)}$$

where $\delta\nu_Q(0) = \nu_Q S_{C-D}$ is the "aligned sample splitting", obtained when the director is parallel to the magnetic field.

The NMR spectrum of a heliconical nematic with "tilt" angle $\theta_0$ will be simply the uniform superposition of line-shapes of the form $\Lambda\left(\nu; \pm\delta\nu_Q(\theta_0,\varphi), w(\theta_0,\varphi)\right)$ for all possible angles $0 < \varphi < 2\pi$,

$$f_{90-flip}(\nu) = \frac{1}{2\pi}\int_0^{2\pi} d\varphi \sum_{r=\pm 1} \left( \begin{array}{l} \Lambda\left(\nu; r\frac{\delta\nu_Q(0)}{4}\left(3\left(\sin\theta_0 \sin\varphi\right)^2 - 1\right), w(\varphi)\right) + \\ \Lambda\left(\nu; r\frac{\delta\nu'_Q(0)}{4}\left(3\left(\sin\theta_0 \sin\varphi\right)^2 - 1\right), w'(\varphi)\right) \end{array} \right) \quad \text{(S.7)}$$

In eq. (S.7) the parameters $\delta\nu_Q(0)$ and $\delta\nu'_Q(0)$ are the experimental "aligned sample splitting". For the angular dependence of the line-widths we have used the standard form (see refs. [2]) $w(\hat{H}\cdot\hat{n}) = w(0) + 2\left(w(0) - w(90)\right)\left(|P_2(\hat{H}\cdot\hat{n})| - 1\right)$ and similarly for $w'(\hat{H}\cdot\hat{n})$ , with the values of $w(0), w(90), w'(0), w'(90)$ taken from Table I. The calculated NMR spectra using eq (S.7) for various values of the heliconical angle $\theta_0$ are presented in figure S3. These results demonstrate clearly that when $\theta_0 = 0$ (no heliconical structure) the calculated spectrum coincides practically with the experimental spectrum while a qualitative deviation becomes apparent by simple visual inspection already from $\theta_0 = 10^0$ . The deviation from the measured spectral patterns become rapidly more pronounced, leading to a totally different spectrum, for lager values of $\theta_0$.

[2] P. J. Collings, D. J. Photinos, P. J. Bos, P. Ukleja, and J. W. Doane, Phys. Rev. Lett. **42**, 996 (1979).
D. J. Photinos, P. J. Bos, J. W. Doane, and M. E. Neubert, Phys. Rev. A **20**, 2203 (1979).

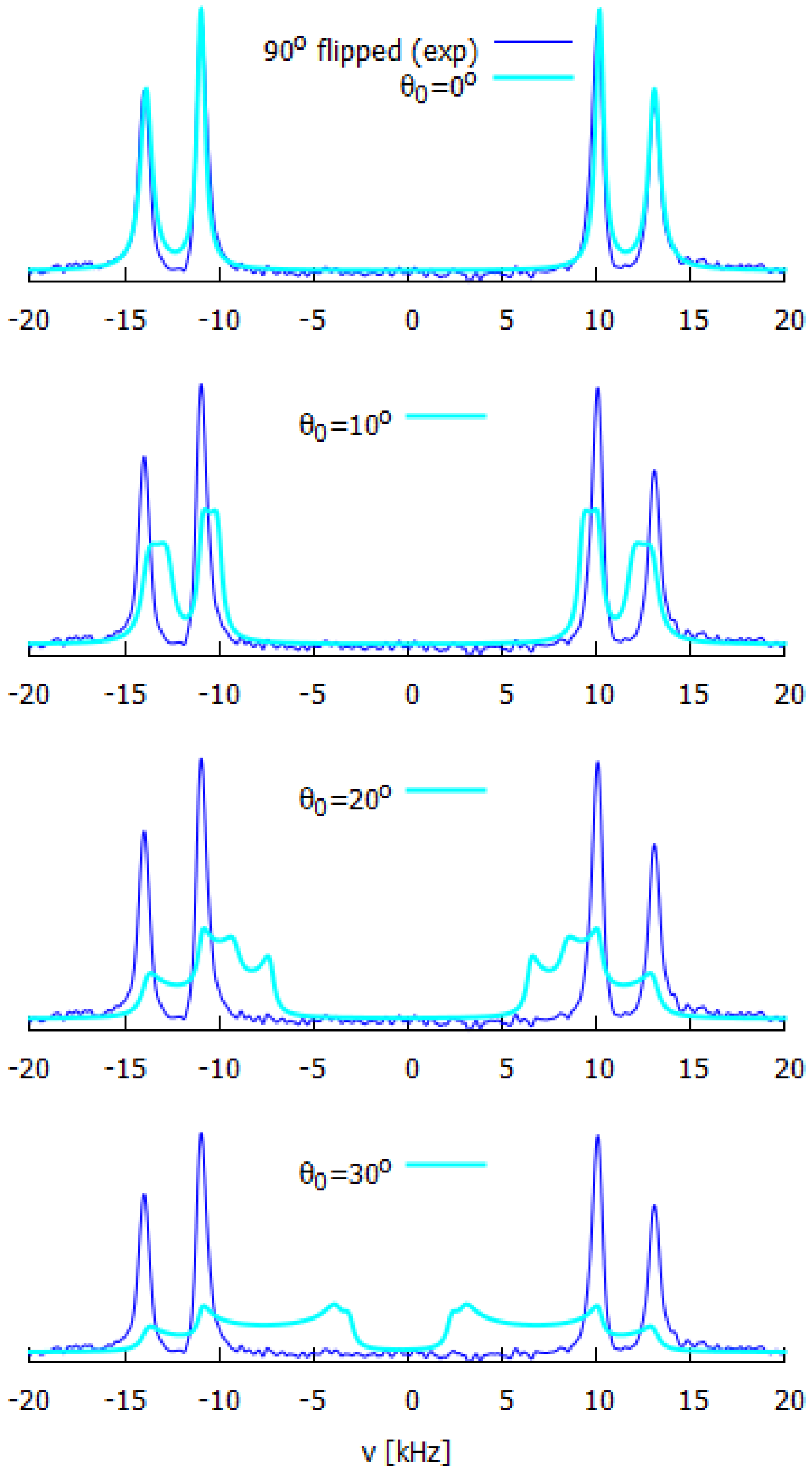

**Figure 3**. NMR spectrum of the 90o-flipped sample (blue) and the corresponding calculated spectra (cyan) for a single, uniformly distributed, director ($\theta_0 = 0$) and for a hypothetical heliconical distribution of the director, $\mathbf{n} = (\sin\theta_0 \cos kZ, \sin\theta_0 \sin kZ, \cos\theta_0)$, with values of the conical angle $\theta_0 = 10^\circ, 20^\circ, 30^\circ$.

It is clear from equations S.5 to 7 that the measured NMR spectrum can pick up variations of the angle of the director relative to the spectrometer magnetic field and that these variations are manifested as broadening of the spectral lines, in excess to their width at perfect alignment. The resolution of the experimental method sets a lower bound to the extent of variations that can be detected unambiguously on the measured spectra. With the simulation procedure described above we have found that, within the resolution of the experimental spectra, both the aligned sample spectrum and the $90^0$-flipped spectrum place an upper bound of $5^0$ for possible variations of the angle of the director relative to the magnetic field. Accordingly, if it is assumed that, as a result of a hypothetical heliconical distribution, the director forms a constant angle $\theta_0$ with the magnetic field, then it is directly deduced from the $90^0$-flipped spectrum that this angle should be below $5^0$. Of course variations of such a limited range could also be due to minor fluctuations of the director within the sample, not related to any heliconical distribution. In any case, it should be noted that the upper bound for $\theta_0<5^0$, set by the combined experimental and simulation resolution, is far below the values given for $\theta_0$ in the literature[3] on the twist-bend nematic phase.

### 3.2 Order parameters of the mesogenic units

The relatively low values implied by the NMR measurements for the order parameter $S_{core}$ of the mesogenic units are explained in the main text in terms of the lack of colinearity of the two mesogenic units in a statistically significant set of conformations of the odd-spacer dimers. An entirely different explanation is provided within the heliconical model of the director configuration. There, the mesogenic units are taken to be well aligned about the director and the reduced values obtained in the aligned sample measurements are attributed to the constant angle $\theta_0$ that the director makes with the magnetic field. Thus, for example, using the relation $\Delta\nu = -\frac{3}{2} q_{CD} S_{a-CD} P_2(\hat{H}\cdot\hat{n})$ for the general case when the local nematic director is not along the magnetic field, and inserting the suggested[4] value $\theta_0 \approx 25°$, we obtain for the order parameter $S_{core}$ of the mesogenic units at 90°C, where the average splitting is $\overline{\Delta\nu} \approx 49kHz$, the value $S_{core} \approx 0.8$. This is a rather high value, particularly in view of the fact that the value of $S_{core}$ for the same compound in the crystal phase[5] is considerably lower than 1. Furthermore, on adding a few more degrees to the used value of $\theta_0$, according to other suggestions in the literature (see ref 3(e,f) above) the resulting values of $S_{core}$ would easily exceed the physical bound of 1, signaling most directly a serious inconsistency of the heliconical interpretation of the order parameter measurements.

[3] (a) V. Borshch, Y.-K. Kim, J. Xiang, M. Gao, A. Jákli, V. P. Panov, J. K. Vij, C. T. Imrie, M. G. Tamba, G. H. Mehl, and O. D. Lavrentovich, Nat. Commun. **4**, 2635 (2013).
(b) D. Chen, J. H. Porada, J. B. Hooper, A. Klittnick, Y. Shen, M. R. Tuchband, E. Korblova, D. Bedrov, D. M. Walba, M. A. Glaser, J. E. Maclennan, and N. A. Clark, PNAS **110**, 15931 (2013).
(c) L. Beguin, J. W. Emsley, M. Lelli, A. Lesage, G. R. Luckhurst, B. A. Timimi, and H. Zimmermann, J. Phys. Chem. B **116**, 7940 (2012).
(d) C. Meyer, G. R. Luckhurst, and I. Dozov, Phys. Rev. Lett. **111**, 067801 (2013).
(e) J. W. Emsley, M. Lelli, A. Lesage, and G. R. Luckhurst, J. Phys. Chem. B **117**, 6547 (2013).
(f) C. Greco, G. R. Luckhurst, and A. Ferrarini, Phys. Chem. Chem. Phys. **15**, 14961 (2013).

**3.3 Viscosity and molecular diffusion in the $N_X$ phase.**

Attempts to reconcile the possibility of a heliconical configuration of the director with the presence of a single direction of orientational averaging implied by the NMR measurements often invoke the possibility of molecular diffusion. Specifically, fast (on the NMR time-scale) molecular diffusion would lead to the sampling of a large enough set of director orientations, in the presumably heliconical arrangement, to make the spectrum correspond to an effective orientational averaging about the direction of the heliconical axis. However, this interpretation can be dismissed since the possibility of such fast diffusion over distances covering the full range of different orientations of the macroscopic director is not supported by the very high viscosity of the Nx phase. Notably, it is precisely such elevated viscosity that gives rise to the long rotational relaxation times allowing the acquisition of the rotated spectra in the Nx phase and not in the lower viscosity normal N phase of the same compound (see section 2). Accordingly, the mechanism of fast molecular diffusion could be applicable in the case of the N phase rather than the $N_X$.

[4] D. Chen, J. H. Porada, J. B. Hooper, A. Klittnick, Y. Shen, M. R. Tuchband, E. Korblova, D. Bedrov, D. M. Walba, M. A. Glaser, J. E. Maclennan, and N. A. Clark, Proc. Natl. Acad. Sci. **110**, 15931 (2013).
[5] K. Hori, M. Iimuro, A. Nakao, and H. Toriumi, J. Mol. Struct. **699**, 23 (2004).

## 4. DSC Trace of CB-C9-CB at 10°C min$^{-1}$

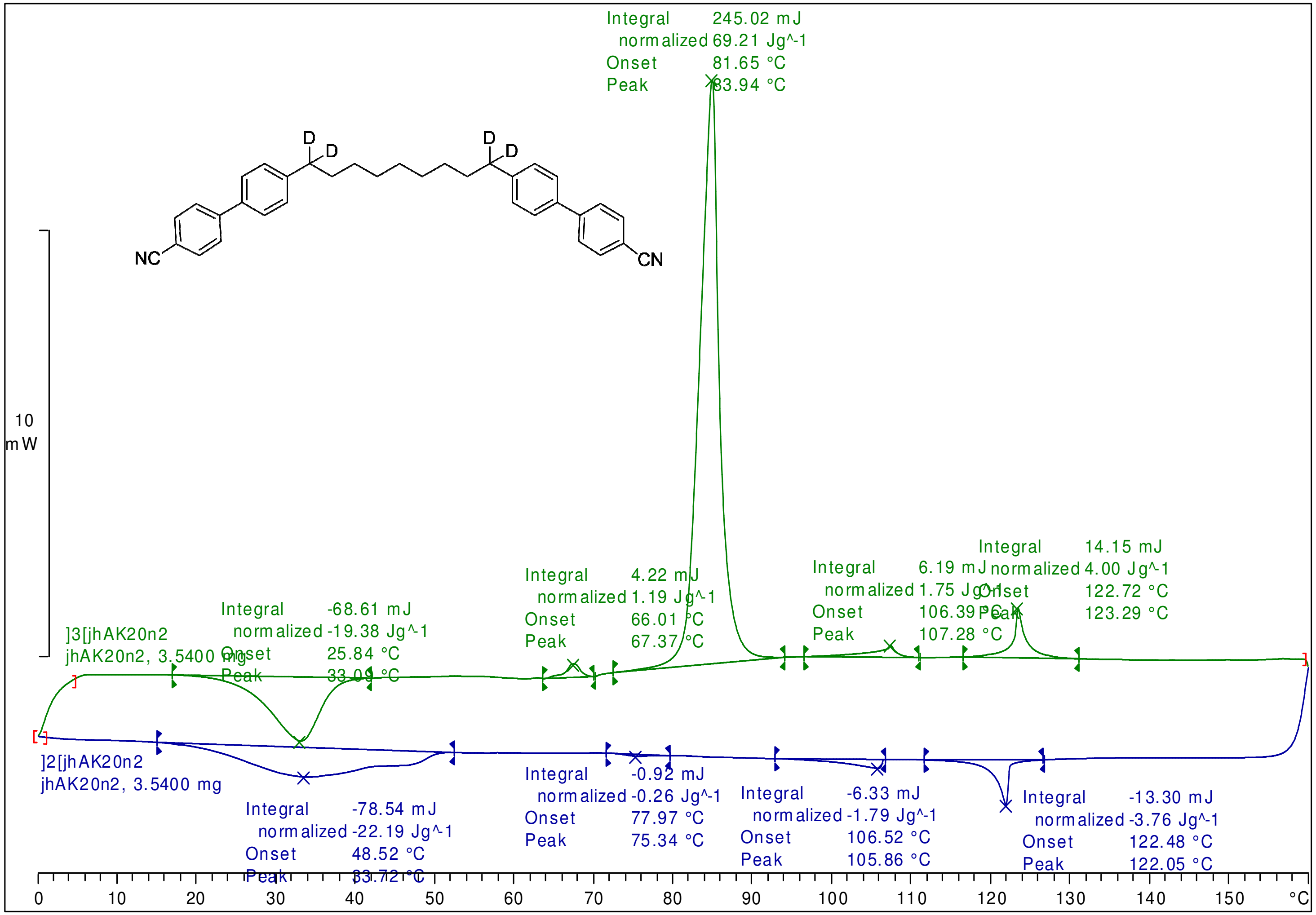